\documentclass[journal,10pt]{IEEEtran}

\pdfoutput=1

\usepackage{graphicx}
\usepackage[cmex10]{amsmath}
\usepackage[utf8]{inputenc}
\usepackage{epstopdf}
\usepackage{graphicx}
\usepackage{epstopdf}
\usepackage{array}
\usepackage{floatrow}
\usepackage{caption}
\usepackage{subcaption}
\usepackage{gensymb}
\usepackage{url}
\usepackage{balance}
\usepackage{enumerate} 
\usepackage{amsbsy}
\usepackage[normalem]{ulem}
\usepackage{textcomp}
\usepackage{siunitx}
\usepackage{balance}
\usepackage{placeins}
\usepackage{tabularx, ragged2e}
\usepackage{listings}
\usepackage[dvipsnames]{xcolor}
\usepackage{svg}
\usepackage{amssymb}
\usepackage{stfloats}
\usepackage{subcaption}
\usepackage{booktabs}
\usepackage{multirow}
\usepackage{adjustbox}
\usepackage{tabularx}
\usepackage{pifont}
\usepackage{makecell}

\usepackage{hyperref}

\usepackage[compact]{titlesec}
\titlespacing*{\subsection}{0pt}{0.8ex plus .2ex}{0.3ex}

\usepackage{cleveref}

\def\BibTeX{{\rm B\kern-.05em{\sc i\kern-.025em b}\kern-.08em
    T\kern-.1667em\lower.7ex\hbox{E}\kern-.125emX}}
    
\usepackage{xspace}

\DeclareTextFontCommand{\textcomputer}{\fontfamily{cmr}\selectfont}

\newcommand{\nicepar}[1]{\smallskip \noindent \textbf{#1}}

\newcommand{\methname}{CAMASA}

\begin{document}

\bstctlcite{IEEEexample:BSTcontrol}

\title{\methname{}: A CAM-based Dataset from the\\MASA Living Lab}

\author{\IEEEauthorblockN{Salvatore~Iandolo, Marco~Savarese, Gaetano~Orazio~Cauchi, Antonio~Solida,\\ Martin~Klapez, Maurizio~Casoni, Angelo~Porrello and Carlo~Augusto~Grazia}\\
\IEEEauthorblockA{\textit{Department of Engineering ``Enzo Ferrari''},
\textit{University of Modena and Reggio Emilia}\\
\{name.surname\}@unimore.it}\thanks{The work of S.~Iandolo, M.~Savarese, G. O.~Cauchi, A.~Solida, M.~Klapez, M.~Casoni, A.~Porrello and C. A.~Grazia was carried out within the MOST – Sustainable Mobility National Research Center and received funding from the European Union Next-GenerationEU (PIANO NAZIONALE DI RIPRESA E RESILIENZA (PNRR) – MISSIONE 4 COMPONENTE 2, INVESTIMENTO 1.4 – D.D. 1033 17/06/2022, CN00000023).}
}

\maketitle
\thispagestyle{empty}  %

\begin{abstract}
Trajectory prediction is a key enabler of autonomous and cooperative driving systems. However, most existing benchmarks are either sensor-centric, geographically constrained, or based on synthetic mobility traces that do not capture real-world V2X communication dynamics.
This paper introduces \methname{}, a large-scale infrastructure-based dataset derived from Cooperative Awareness Messages (CAMs) and Decentralized Environmental Notification Messages (DENMs) collected within the Modena Automotive Smart Area (MASA). The dataset comprises more than 40 million CAMs and 2 million DENMs recorded under authentic urban traffic conditions over multiple months.
We present a rigorous preprocessing pipeline that includes filtering, pseudonym reconciliation to account for ETSI privacy-driven stationID changes, and temporal normalization to 10 Hz trajectories, suitable for motion forecasting and time-series analysis. With over 14,000 km of reconstructed vehicle paths and tens of thousands of unique station IDs, \methname{} provides a statistically significant empirical foundation for research on Cooperative Intelligent Transportation Systems (C-ITS).
Beyond trajectory prediction, the dataset enables calibration of microscopic urban traffic simulators (e.g., SUMO) and supports the development of realistic Intelligent Transportation Systems (ITS) Digital Twins by jointly modeling mobility patterns and V2X communication coverage in real deployments.
\end{abstract}


\section{Introduction}
\label{sec_intro}

Artificial Intelligence (AI), in addition to being the technology that has come to symbolize recent years of innovation, is itself a generator of new technologies and opportunities. Among its most interesting applications, its use in the automotive sector is undoubtedly prominent, with particular focus on remote and autonomous driving. The latter, however, overshadows a broad range of research topics that integrate AI with the automotive world.

\begin{figure}
    \centering
    \includegraphics[width=1\textwidth]{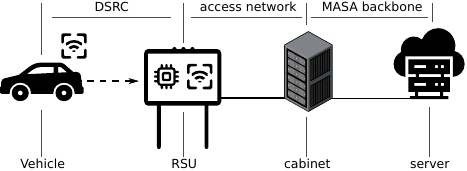}
    \caption{MASA: real living-lab network.}
    \label{fig:masa}
\end{figure}

A representative example is trajectory prediction, arguably the most important driver for autonomous vehicles. This task analyzes the historical movements of objects (\emph{e.g.}, vehicles, pedestrians, \emph{etc.}) and aims to estimate their future paths. Its role is critical for the safety of future road users, as accurate trajectory prediction could enable complete collision avoidance and thereby substantially improve safety. This research area has seen numerous advances in recent years. Among them, ForecastMAE~\cite{cheng2023forecast} is a notable contribution that employs masked autoencoders to learn spatiotemporal representations of agent trajectories, achieving state-of-the-art performance in multi-modal prediction scenarios. Similarly, the work most closely related to this one, CAMNet~\cite{grasselli2026camnet}, addresses the trajectory prediction problem by combining a Variational Autoencoder (VAE), a Recurrent Neural Network (RNN), and a Graph Neural Network.

Since trajectory prediction relies on historical trajectories to predict future ones, the foundation of such systems lies in their ability to access accurate historical data. The most traditional approaches to recover historical data rely on onboard sensor fusion, which combines data from different sources, such as LiDARs, cameras, and radars, like in Argoverse 2~\cite{wilson2023argoverse}. However, this approach presents inherent limitations: perception range is constrained by sensor specifications and environmental conditions, while occlusions can completely obscure critical information about nearby agents. Furthermore, vehicles can only perceive what their onboard sensors directly observe, creating blind spots that may compromise prediction accuracy in dense traffic scenarios.

An alternative approach, already validated in the CAMNet paper, leverages Cooperative Awareness Messages (CAMs) and Decentralized Environmental Notification Messages (DENMs), cornerstones of Vehicle-to-Everything (V2X) communication within the European Telecommunications Standards Institute (ETSI) Intelligent Transportation Systems (ITS) framework. Particular attention will be paid to CAMs, which are periodically broadcast by connected vehicles and contain essential information, including vehicle position, speed, acceleration, and heading. This contrasts with DENMs, which are event and situation-triggered and issued in response to specific road hazards or events. By aggregating CAMs from multiple vehicles, CAMNet constructs a comprehensive, shared situational awareness that extends far beyond the line-of-sight limitations of traditional sensors. This cooperative paradigm not only provides richer contextual information but also enables trajectory prediction in scenarios where vehicles may be occluded from a particular observer's perspective, ultimately contributing to the overarching goal of collision-free autonomous navigation. CAM and DENM aggregation was enabled by the Roadside Units (RSUs) of the Modena Automotive Smart Area (MASA), which will be described later in this article.
Beyond trajectory prediction, large-scale collections of standardized V2X messages also provide a valuable empirical basis for traffic modeling and communication-aware mobility analysis. Real-world CAM traces enable the calibration of microscopic traffic simulators using observed kinematic behaviors and offer quantitative insights into the interaction between traffic dynamics and radio coverage conditions in urban deployments. Such datasets are therefore not only relevant for forecasting tasks, but also for the development of realistic ITS Digital Twins grounded in operational data rather than purely synthetic traces.

Given this context, this paper introduces \methname{}, a large-scale CAM/DENM dataset designed to support communication-aware mobility and protocol-level analysis, trajectory modeling, and empirical validation of Cooperative ITS (C-ITS) systems. Beyond addressing the dataset characterization gap left by CAMNet, this work aims to provide an empirically grounded real-world foundation for both data-driven forecasting research and simulation-based ITS evaluation. The paper presents a detailed quantitative analysis of the dataset, a comprehensive description of the filtering and preprocessing pipeline applied to the raw messages, and a structured comparison with existing literature. To summarize, the main contributions of this work are:

\begin{itemize}
    \item The release of one of the largest real-world ETSI-compliant CAM/DENM datasets collected in an urban living lab, enabling statistically grounded traffic and communication analysis.
    \item A detailed characterization of MASA as a large-scale urban C-ITS experimental infrastructure.
    \item A pseudonym-reconciliation pipeline enabling long-term trajectory reconstruction across station ID changes.
    \item A temporally normalized 10Hz processed release tailored for trajectory prediction, motion forecasting, and time-series mobility analysis.
\end{itemize}

The remainder of this paper is organized as follows.
\Cref{sec_related} provides an overview of related works.
\Cref{sec_masa} introduces MASA and details its sensing and computing architectures.
\Cref{sec_dataset} illustrates the dataset, specifically the operations that translate raw messages into structured, usable data, and describes its contents and statistical properties. 
\Cref{sec_conclusions} concludes the article.

The CAMASA dataset is publicly available for download at \url{https://www.automotivesmartarea.it/dataset/}.

\section{Related Works}
\label{sec_related}

\begin{figure}
    \centering
    \includegraphics[width=1\textwidth]{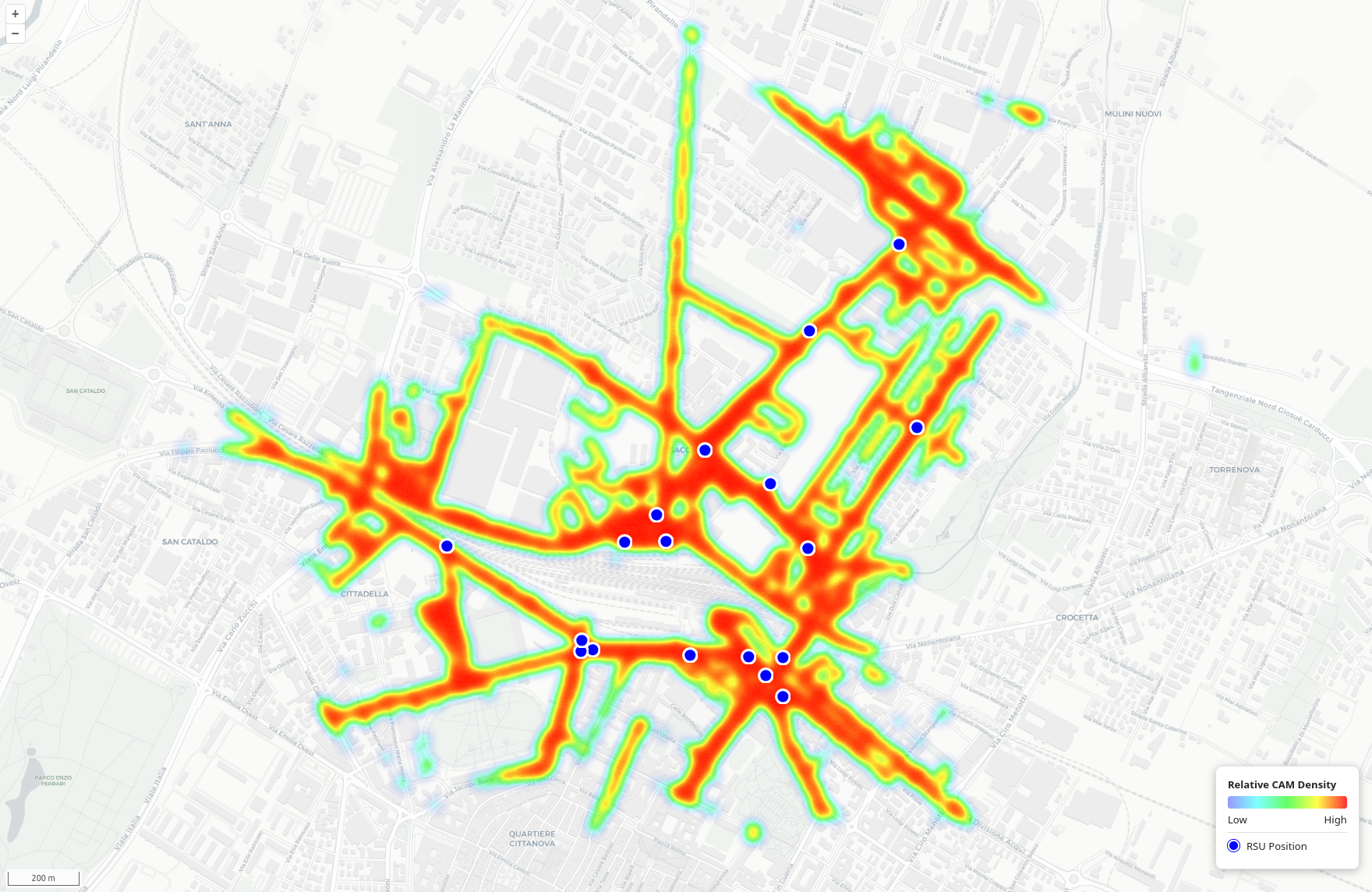}
    \caption{CAM density map MASA}
    \label{fig:cam_density}
\end{figure}

To the best of our knowledge, no publicly available dataset matches the scale, temporal continuity, or message density of the dataset presented in this work. Additionally, no publicly available CAM dataset provides a pseudonym-reconciled trajectory-level release. Existing contributions typically focus on smaller-scale deployments, shorter acquisition windows, synthetic mobility traces, or perception-oriented data, thereby limiting their suitability for large-scale, long-term evaluation of standardized C-ITS communication under sustained real-world operation.

In autonomous driving, significant progress has been achieved through large-scale datasets such as Argoverse 2~\cite{wilson2023argoverse}. This suite represents a major evolution by offering three distinct datasets tailored for perception and motion forecasting across six diverse U.S. cities (Austin, Detroit, Miami, Palo Alto, Pittsburgh, and Washington D.C.). It comprises $1,000$ multimodal sequences with high-resolution imagery and 3D annotations across $30$ object categories, along with the largest collection of unlabeled LiDAR sequences ($20,000$) for self-supervised learning. Crucially, all three components are anchored by rich HD maps with 3D lane and crosswalk geometry. While Argoverse 2 excels in geographic diversity and sensor fusion, its primary focus remains on perception and trajectory prediction rather than standardized V2X communication logs.

Moving toward real-world vehicular communications, the V2AIX dataset~\cite{kueppers2024v2aix} marks a milestone by providing over $285,000$ recorded ETSI ITS messages (primarily CAM) collected from more than $2,380$ vehicles and RSUs. Unlike earlier works that rely on synthetic traces, V2AIX provides a valuable empirical basis for analyzing standardized V2X protocols. A key contribution is its integration of ETSI message formats into the Robot Operating System (ROS), enabling interoperability between communication stacks and automated driving frameworks. However, despite its methodological scope, the overall message volume and observation horizon remain limited when compared to large-scale, long-term monitoring campaigns.

To address the fragmentation of available vehicular datasets, Bari \emph{et al.}~\cite{bari2025datasets} provided a comprehensive overview, systematically categorizing them by application domain, sensing modality, and technical characteristics. Their analysis highlights persistent limitations in scalability, data representativeness, and multi-modal integration across the field. Specifically, they emphasize that many existing datasets lack extended temporal coverage and real-world heterogeneity—factors essential for validating next-generation ITS protocols under realistic operational stress. Their findings underscore the critical need for large-scale, standardized, and longitudinal datasets to support statistically robust evaluations.

Infrastructure-based sensing offers another avenue for empirical evidence, exemplified by the PASMO project~\cite{ferreira2022dataset}. This initiative introduced a two-year longitudinal dataset derived from a dense network of traffic-classification radars in the Aveiro region. Comprising $74,305$ aggregated records at 10-minute resolution, PASMO captures vehicle speed, position, and class information with strong temporal continuity. Its strength lies in correlating mobility data with exogenous variables, such as meteorological conditions and pandemic-related shifts. Nevertheless, because PASMO relies solely on infrastructure sensors, it does not include standardized V2X communication messages and therefore cannot directly support protocol-level evaluation of ETSI ITS services.

Similarly, the Warrigal Dataset~\cite{ward2014warrigal} provides long-term trajectory and V2V communication logs collected over three years in a quarry environment. While it includes detailed state information and $1Hz$ communication data from $13$ vehicles, its scope is strictly limited to a controlled operational setting. This constraint on fleet size and environment significantly constrains its applicability to large-scale urban C-ITS validation.

From a simulation perspective, researchers have developed tools to bridge the gap between raw data and realistic scenarios. Uppoor \emph{et al.}~\cite{uppoor2013generation} introduced the TAPASCologne dataset, covering a $400Km$ urban area over $24$ hours. By combining real-world travel demand estimation with microscopic traffic simulation in SUMO, this approach addresses inconsistencies in raw mobility data through a structured repair process. Although TAPASCologne significantly improved realism in simulated vehicular networking evaluations, it remains a synthetic trace and does not capture real-world communication channel dynamics or deployment-specific artifacts.

More recently, efforts have focused on cooperative perception benchmarks. Xu \emph{et al.}~\cite{xu2022opv2v} introduced OPV2V, the first comprehensive benchmark for Vehicle-to-Vehicle (V2V) communication. Collected using the CARLA simulator and the OpenCDA framework, OPV2V comprises $73$ diverse scenes and $11,464$ frames featuring over $230,000$ annotated 3D vehicle bounding boxes. Spanning $8$ different CARLA towns and a digital twin of Culver City, Los Angeles, it provides a robust environment for evaluating domain adaptation. Each frame includes synchronized data from $64$-channel LiDARs and $360\degree$ RGB cameras, enabling rigorous benchmarking of fusion strategies under realistic traffic densities.

The push for geographic generalization has also led to the development of diverse datasets, such as the one proposed by Sun \emph{et al.}~\cite{sun2020scalability}. Consisting of $1,150$ twenty-second scenes integrating LiDAR and high-resolution camera data, this work stands out for its $15$-fold higher environmental diversity compared to existing benchmarks. By providing consistent IDs over time, it establishes robust baselines for tracking and detection while critically analyzing the impact of dataset scalability on cross-geographic generalization capabilities.

Finally, perception-oriented datasets such as the Indian Traffic Dataset (ITD)~\cite{agarwal2024itd} address visual detection challenges under heterogeneous traffic conditions. ITD includes over $30,000$ annotated images across $13$ vehicle categories, evaluating modern deep learning architectures under complex, mixed-traffic scenarios. While highly relevant for the computer-vision components of ITS, such datasets do not provide standardized V2X communication logs and therefore address a complementary, rather than overlapping, research dimension.

In summary, while the literature offers valuable insights into perception, simulation, and specific communication scenarios, none of these resources simultaneously satisfies the requirements for large-scale, long-term temporal continuity, high message density, and pseudonym-reconciled trajectory-level data necessary for a holistic evaluation of C-ITS systems.

\begin{table}
    \centering
    \renewcommand{\arraystretch}{1.0}
    \resizebox{\textwidth}{!}{%
    \begin{tabular}{lccccc}
    \toprule
    &\multicolumn{2}{c}{\# of messages}&\multicolumn{2}{c}{Unique IDs}\\
    \textbf{RSU IP} & CAM & DENM & CAM & DENM & Distance ($Km$)\\
    \midrule
    \textbf{12.121} & $3,290,911$ & $210,708$ & $11,399$ & $956$& $9,042$\\
    \textbf{12.122} & $3,096,359$ & $184,796$ & $9,694$ & $869$& $8,463$\\
    \textbf{12.123}& $3,317,357$ & $241,025$ & $11,614$ & $1,054$ & $10,179$\\ 
    \midrule
    \textbf{Total} & $40,446,000$ & $2,376,571$ & $25,839$ & $1,332$ & $103,054$\\
    \bottomrule
    \end{tabular}
    }
    \caption{Selected RSUs statistics.}
    \label{tab:RSU_stats}
\end{table}
\begin{table}
    \centering
    \renewcommand{\arraystretch}{1.0}
    \resizebox{\textwidth}{!}{%
    \begin{tabular}{lccc}
    \toprule
    \textbf{CauseCode | subCauseCode} & Unique IDs & Total Messages\\
    \midrule
    \textbf{1 | 0 (Traffic Increasing)}  & $1,106$ & $1,606,678 $\\
    \textbf{94 | 0 (Stationary vehicle)}  & $198$ & $767,711$\\
    \textbf{99 | 0 (Dangerous situation)}  & $20$ & $2,182$\\
    \bottomrule
    \end{tabular}
    }
    \caption{DENM statistics.} 
    \label{tab:DENM_stats}
\end{table}

\section{MASA in a nutshell}
\label{sec_masa}
The Modena Automotive Smart Area (MASA,\footnote{https://www.automotivesmartarea.it/}~\Cref{fig:masa}) is an open-air urban test facility conceived to enable experimentation, validation, and pre-deployment assessment of connected and cooperative mobility solutions under realistic traffic conditions. The infrastructure extends across selected areas of the city of Modena and the university campus, covering a total surface area of approximately $0.64Km^2$. This extensive configuration supports comprehensive testing of ITS and V2X services in operational environments.

The test site is equipped with a dense deployment of $18$ ETSI-compliant RSUs, arranged to ensure pervasive V2X coverage. These units are interconnected through a high-capacity optical fiber backbone that forms the MASA core network, thereby guaranteeing low-latency, reliable connectivity among roadside devices, edge computing platforms, and centralized backend systems. Such an architecture enables hybrid communication schemes in which delay-sensitive information is exchanged via short-range wireless links, while computationally intensive processing is delegated to edge or cloud resources.

The RSUs, produced by Movyon Electronics, comply with the European ITS-G5 (IEEE 802.11p) standard and the C-V2X specifications defined in 3GPP Releases $14$ and $15$. Operating in the regulated $5.9GHz$ ITS band with a reception sensitivity of $-97dBm$, the devices integrate a dedicated ITS-G5 protocol stack and multi-constellation GNSS for accurate geolocation. The hardware is CE RED certified and enclosed in IP67-rated casings for outdoor deployment.

Beyond over-the-air V2X communications, MASA supports synchronized acquisition of both vehicular and infrastructure data streams. In particular, the platform allows monitoring and recording of ETSI ITS messages, namely CAMs and DENMs. The combined logging of CAM data and DENM events facilitates cross-layer performance analysis of cooperative safety applications, thereby supporting evaluation of latency, reliability, and event-dissemination effectiveness in real-world traffic scenarios.

\section{Dataset}
\label{sec_dataset}

The dataset comprises large-scale ITS messages collected via V2X communications from real production vehicles, primarily Volkswagen models equipped with factory-installed V2X stacks. The dataset contains more than $40$ million CAMs and $2$ million DENMs, collected in a real-world urban scenario and therefore reflecting authentic traffic dynamics rather than simulated mobility patterns.

CAM messages provide detailed information about the transmitting vehicle, including geographic position (latitude and longitude), instantaneous speed, heading, steering angle, pedal activation status (\emph{e.g.}, braking), vehicle dimensions and classification, and precise timestamps. This set of dynamic and static attributes enables fine-grained reconstruction of vehicle trajectories while they traverse the RSU coverage area, allowing infrastructure-based tracking and mobility analysis without relying on onboard data logging. In contrast, DENM messages are event-driven and are transmitted when specific hazardous or abnormal conditions occur. Within this dataset, the identified DENM categories are mainly three: stationary vehicle, traffic jam increasing, and generic hazardous situation. These safety-related notifications can be correlated with CAM-based kinematic data to support integrated analyses of traffic conditions and risk scenarios.

\Cref{tab:RSU_stats} reports the quantitative details of the collected data, including the total number of CAM and DENM messages, the three RSUs with the highest number of received messages, the number of unique station IDs (\emph{i.e.}, uniquely identified vehicles) tracked during the observation period, and the cumulative distance (expressed in kilometers) reconstructed from the detected vehicle trajectories. These aggregated indicators provide an overall view of the dataset scale, infrastructure load distribution, vehicle population coverage, and the spatial extent of the monitored mobility.

\Cref{tab:DENM_stats} focuses specifically on DENM statistics, detailing the total number of received DENMs, the three RSUs that recorded the highest number of such messages, their distribution across the three identified categories (stationary vehicle, traffic jam increasing, and generic hazardous situation), and the number of unique station IDs associated with DENM generation. This complementary breakdown highlights the spatial concentration and typological distribution of safety-related events within the monitored urban area.

\begin{table*}[t]
    \centering
    \scriptsize
    \resizebox{\textwidth}{!}{%
    \begin{tabular}{lcccccccccc}
        \toprule
        \textbf{stationID} & src\_addr & timestamp & Lon & Lat & Heading & Speed & genDeltaTime & x & y & interp \\
        \midrule
        \textbf{186428156} & -- & 2025-04-19 14:34:19.621  & $10.9249241$ & $44.6608201$ & $119.6$ & $14.47$ & $8,238$ & $652,606.26$ & $4,947,075.19$ & True \\
        \textbf{186428156} & a2:e3:0b:1c:aa:fc & 2025-04-19 14:34:19.720 & $10.9249392$ & $44.6608136$ & $120.5$ & $14.57$ & $8,331$ & $652,607.48$ & $4,947,074.49$ & False \\
        \textbf{186428156} & a2:e3:0b:1c:aa:fc & 2025-04-19 14:35:25.219 & $10.9286462$ & $44.6590978$ & $120.8$ & $0.73$ & $8,317$ & $652,905.88$ & $4,946,890.85$ & False \\
        
        \textbf{186428156} & -- & 2025-04-19 14:35:25.320 & $10.9286466$ & $44.6590976$ & $120.8$ & $0.64$ & $8,417$ & $652,905.91$ & $4,946,890.83$ & True \\
        \hline
    \end{tabular}
    }
    \caption{Example of filtered and interpolated CAM data.}
    \label{tab:cam_data_example}
\end{table*}

\begin{figure}
    \centering
    \includegraphics[width=1\textwidth]{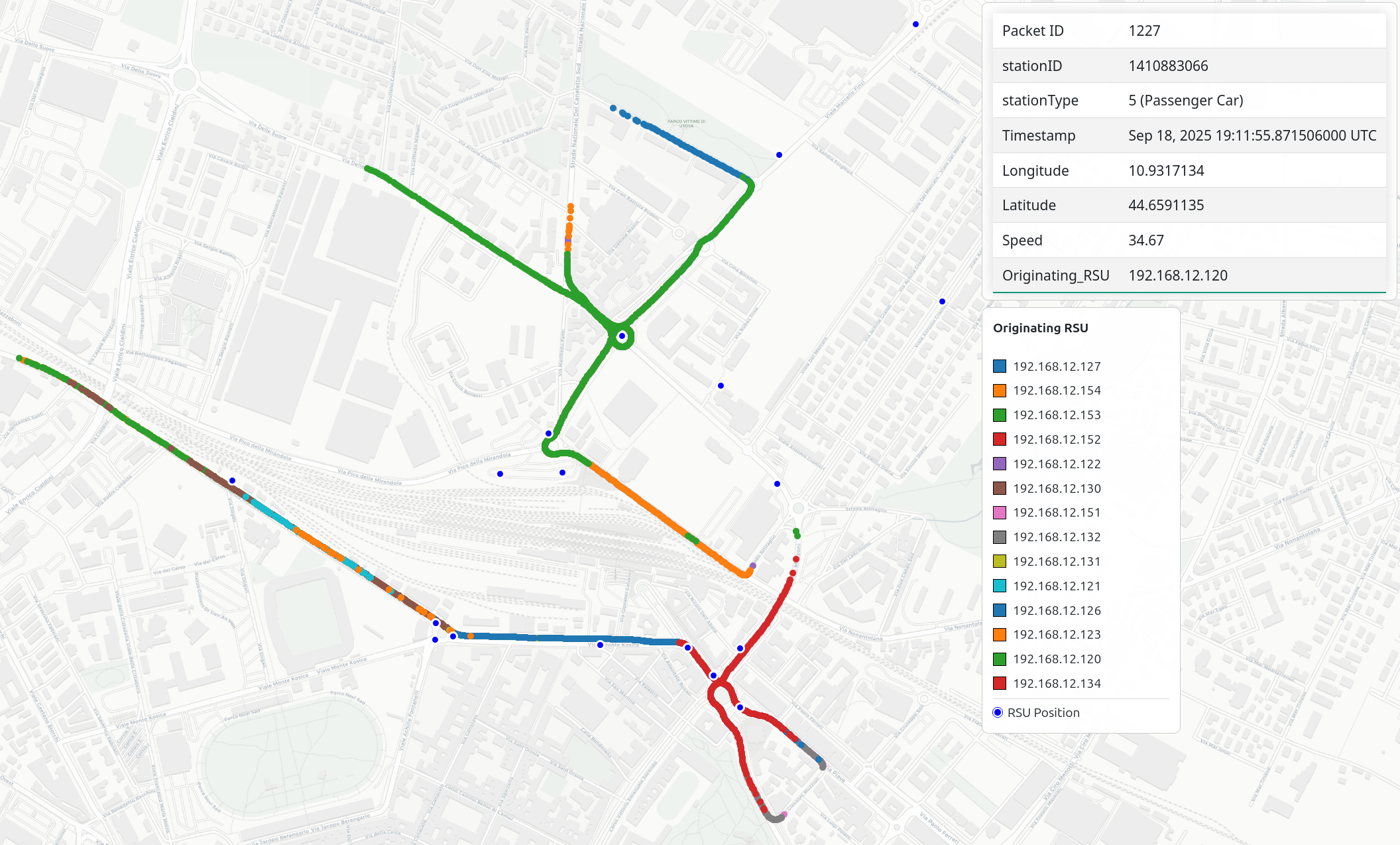}
    \caption{Vehicle trajectory via multiple RSUs.}
    \label{fig:car_trajectory}
\end{figure}

The spatial characteristics of the dataset are shaped by the urban road topology and the RSU deployment strategy within the MASA. Since data collection is entirely infrastructure-based, the spatial distribution of received messages depends on RSU placement, radio coverage conditions, traffic density, and the penetration rate of V2X-enabled vehicles. As vehicles move through the monitored area, their periodic CAM transmissions are captured by one or more RSUs, resulting in a non-uniform spatial density of messages, typically higher along main roads and at busy intersections.
These spatial properties are illustrated in two reference figures. \Cref{fig:cam_density} presents a heat map of the MASA coverage area, highlighting the spatial distribution and frequency of received ITS messages. Higher-intensity areas correspond to zones with denser traffic or stronger overlap in RSU radio coverage, providing insights into both mobility patterns and the effectiveness of infrastructure deployment. 

\Cref{fig:car_trajectory} shows the tracking of a vehicle within the MASA, where the trajectory is reconstructed from CAM-reported positions, and different colors indicate the specific RSU that received each message. The associated table in the figure summarizes key attributes extracted from the messages, such as entity type (\emph{e.g.}, passenger car, motorcycle, pedestrian), speed, timestamp, and other dynamic parameters.

In addition to the raw dataset, we generated a parsed version to facilitate data analysis and trajectory-level studies. In this processed dataset, CAM messages are temporally interpolated to a fixed rate of $10Hz$ to compensate for transmission irregularities and fill potential gaps, given that the observed average transmission interval is approximately $400ms$. The parsed version also includes the extraction of the most relevant attributes (\emph{e.g.}, position, speed, heading, timestamp, and other key kinematic parameters) to provide a cleaner and analysis-ready structure. Furthermore, a vehicle-tracking reconstruction procedure has been implemented to handle station ID changes that occur during a trip due to ETSI privacy mechanisms. This allows the re-association of messages belonging to the same physical vehicle, enabling continuous trajectory reconstruction despite identifier pseudonym changes.

\nicepar{Filtering operations.}~
The filtering operations applied to vehicular packets, specifically the CAMs collected by the Road Side Units (RSUs) of the MASA, were designed to obtain a consistent, traceable, and analysis-ready dataset suitable for quantitative mobility studies. First, the raw PCAP files acquired by the RSUs were parsed and converted to JSON, retaining only the most relevant fields for vehicle tracking (\emph{e.g.}, timestamp, position, speed, heading, and identifiers). This step reduced the data dimensionality and improved the dataset's usability. Subsequently, isolated CAM messages, \emph{i.e.}, single messages not belonging to any temporal sequence, were removed, as they do not allow interpolation or trajectory reconstruction. All GeoNetworking messages transmitted by the RSUs were also discarded, since they are not informative for tracking mobile vehicles.

\begin{figure}
    \subfloat[Segments.]{
    \includegraphics[width=0.47\linewidth]{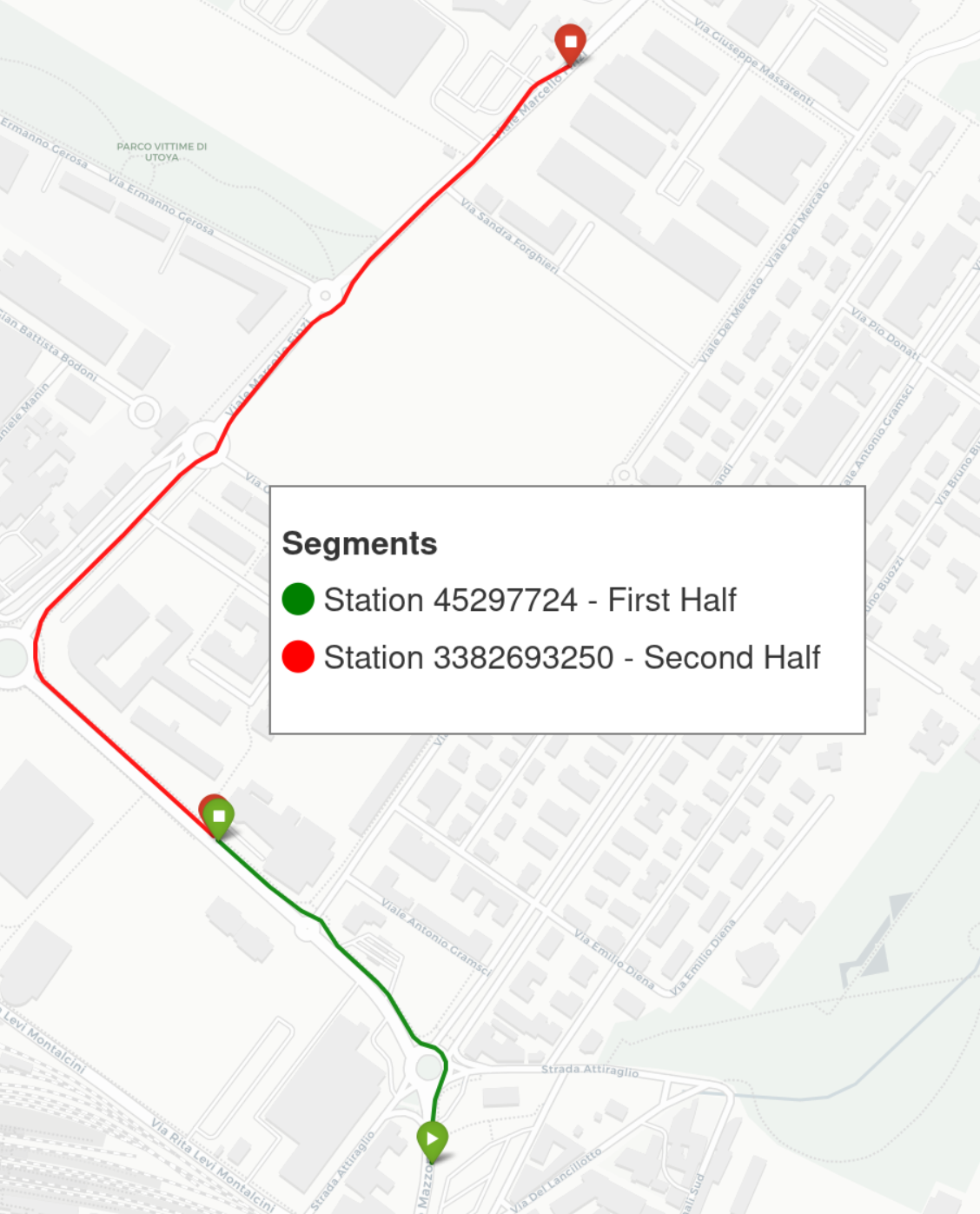}
    }
    \hfill
    \subfloat[Full path.]{
    \includegraphics[width=0.47\linewidth]{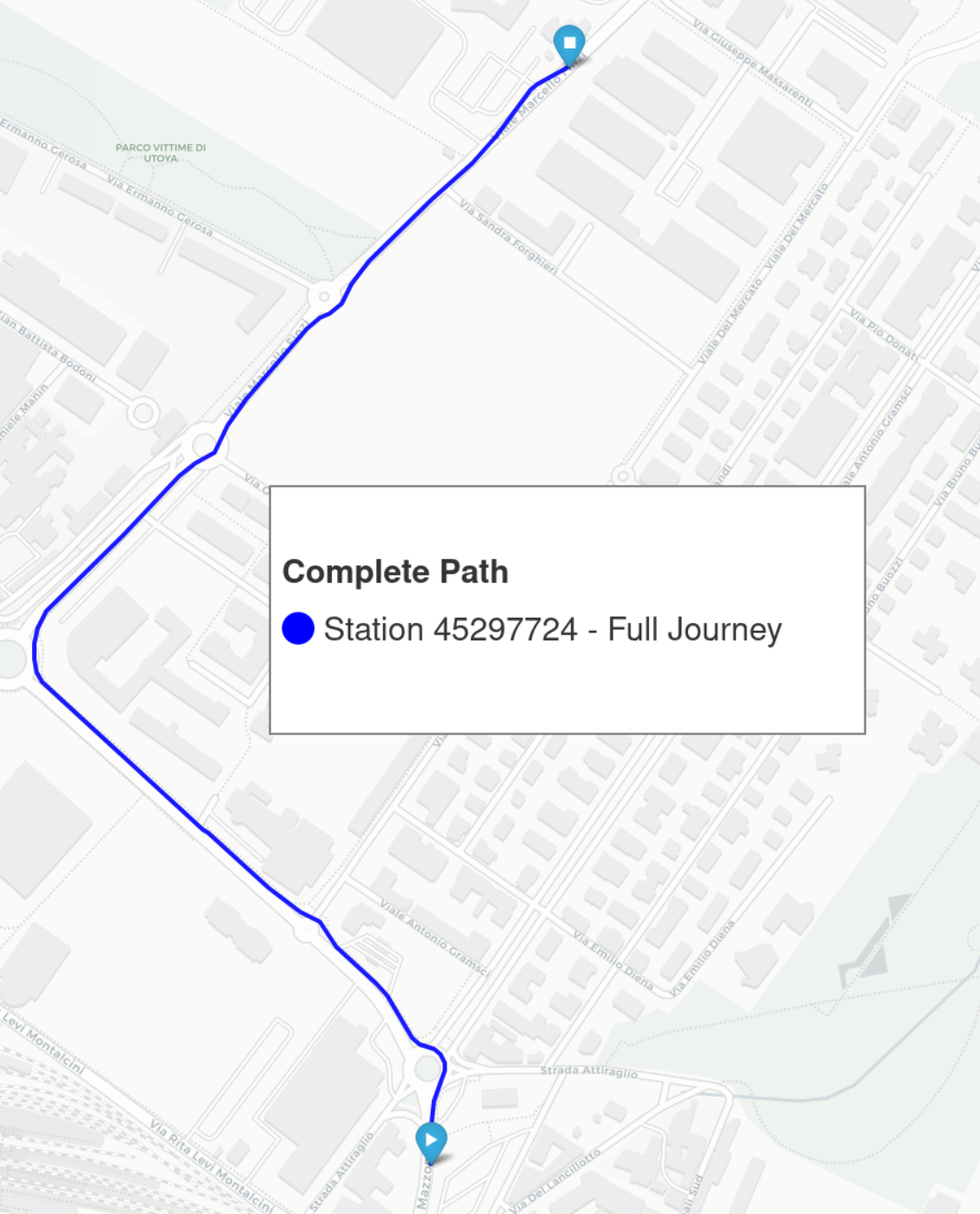}
    }
    \caption{Creation of a single CAM trace.}
    \label{fig:reconciliation}
\end{figure}

An additional cleaning phase excluded specific road segments, particularly ring-road sections characterized by high vehicle speeds and dense traffic, where the trajectory attachment algorithm exhibited degraded performance due to the consecutive passage of vehicles with similar dynamics. All JSON files left empty after the previous filtering steps were then removed from the dataset.

A critical issue is that the privacy-preserving mechanisms defined by ETSI standards allow both the stationID and the MAC address associated with CAM messages to change during vehicle motion, thereby preventing direct, continuous tracking. To address this discontinuity, an iterative reconciliation process was implemented: for each pair of consecutive CAM messages $X$ and $X+1$, if the stationID differs but the temporal gap is equal to $1.5 s$ and the spatial distance does not exceed $21 m$ (corresponding to the maximum distance traveled at $50Km/h$ within the same time interval) the two messages are attributed to the same vehicle. The value of $50km/h$ was selected as it represents the maximum speed limit typically enforced in urban centers. The temporal threshold of $1.5s$ was determined empirically: alternative values within the same range were tested, and $1.5s$ yielded the best reconciliation accuracy in terms of trajectory continuity and reduced false associations. Finally, to obtain uniformly sampled time series suitable for dynamic analyses, linear interpolation (piecewise linear interpolation) was applied to fill the gaps between consecutive messages of the same vehicle, enforcing a constant sampling frequency of $10Hz$. The result of this reconciliation process is illustrated in~\Cref{fig:reconciliation}.

\Cref{tab:cam_data_example} reports a representative excerpt of the dataset obtained after applying all the filtering and pre-processing operations described above. It provides an overview of the structure and content of the processed data used for the analyses presented in this work. For reproducibility and further investigation, the original raw PCAP files are available, enabling additional feature extraction or alternative processing beyond those performed in this study.

\begin{figure}
    \subfloat[Before interpolation.]{
    \includegraphics[width=0.47\linewidth]{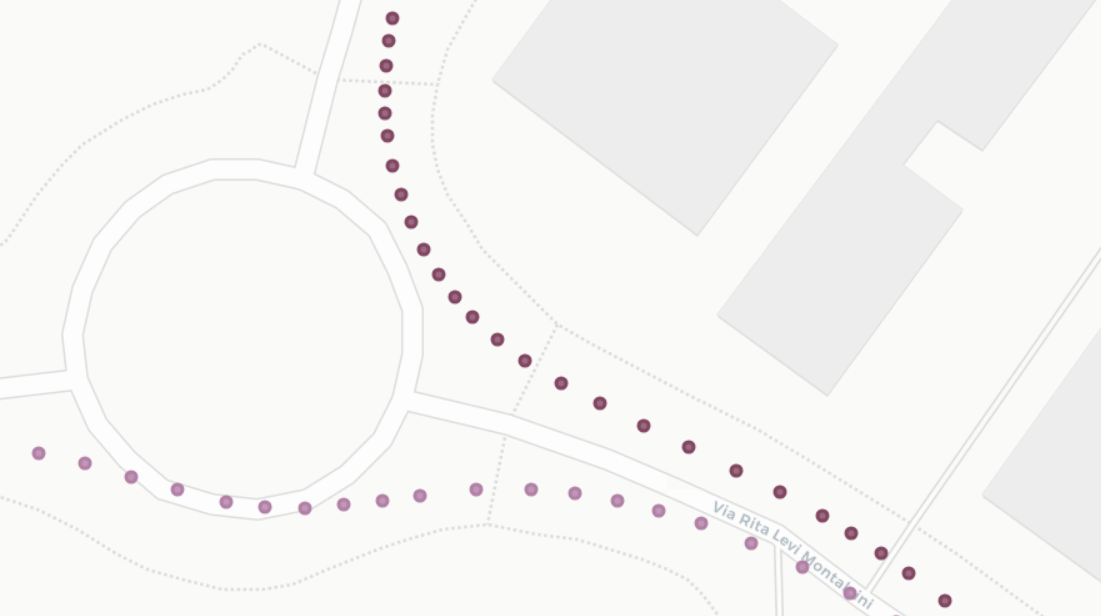}
    }
    \hfill
    \subfloat[After interpolation.]{
    \includegraphics[width=0.47\linewidth]{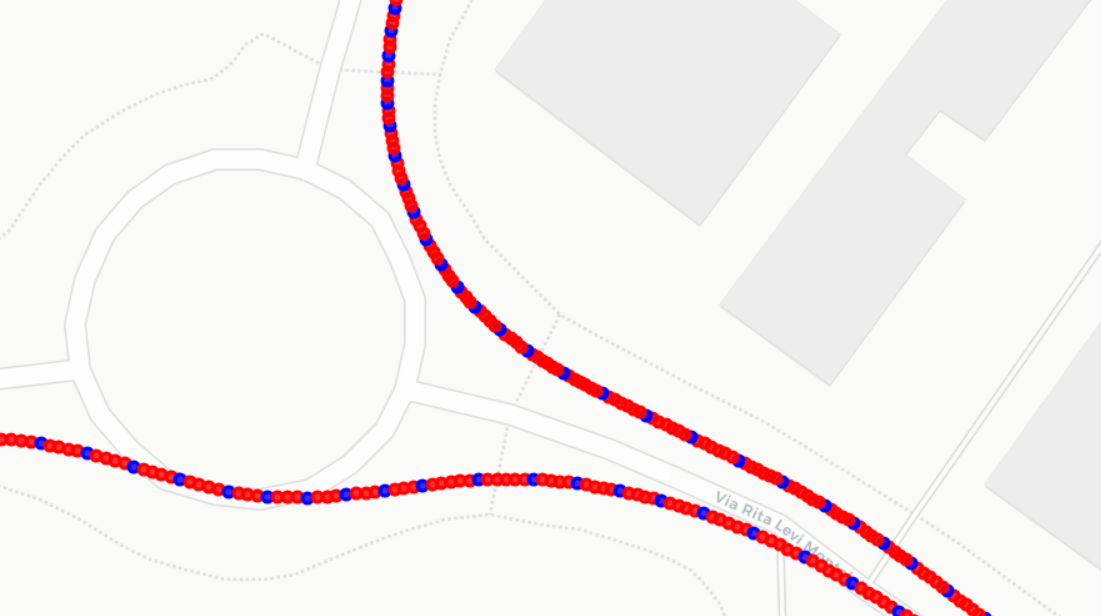}
    }
    \caption{$10Hz$ interpolation process.}
    \label{fig:interpolation}
\end{figure}

\nicepar{Comparison.}~
When compared with the datasets mentioned in~\Cref{sec_related}, \methname{} occupies a complementary position within the C-ITS data ecosystem. Most existing datasets focus either on perception tasks, infrastructure sensing, or simulated mobility traces, and therefore do not provide large-scale collections of standardized ETSI V2X communication messages. For example, PASMO~\cite{ferreira2022dataset} offers long-term infrastructure-based traffic measurements but does not include V2X protocol data, while perception-oriented datasets such as OPV2V~\cite{xu2022opv2v}, the dataset by Sun \emph{et al.}~\cite{sun2020scalability}, and the ITD~\cite{agarwal2024itd} primarily target sensor fusion and visual detection tasks. Similarly, TAPASCologne~\cite{uppoor2013generation} provides realistic mobility traces generated through simulation, but lacks real-world communication dynamics and deployment artifacts. As a result, these datasets address different aspects of ITS research and cannot directly support large-scale empirical evaluation of ETSI cooperative messaging.

Among publicly available datasets, the closest point of comparison is the V2AIX dataset introduced by Kueppers \emph{et al.}~\cite{kueppers2024v2aix}, which also includes standardized ETSI V2X messages collected from real vehicles. Compared to V2AIX, the \methname{} dataset is substantially larger in terms of cooperative awareness data. In particular, the total number of collected CAM messages is nearly two orders of magnitude higher, and the number of DENM messages is also significantly greater. Furthermore, the number of unique station IDs observed in \methname{}—both for CAM and DENM transmissions—is considerably higher, indicating broader coverage of the vehicle population and a larger set of distinct connected vehicles monitored during the observation period.

On the other hand, the V2AIX dataset includes a wider variety of ITS message types. While \methname{} focuses on two message categories (CAM and DENM), V2AIX also incorporates Signal Phase and Timing Extended Messages (SPATEM) and Map Extended Messages (MAPEM), resulting in a total of four ITS message types. Additionally, V2AIX reports a larger set of DENM hazard classes (five compared to the three identified in \methname{}). This difference can likely be attributed to the distinct operational environments in which the datasets were collected. In particular, the V2AIX data collection campaign includes extra-urban and highway scenarios, where certain hazard conditions are more frequently encountered and therefore more easily detectable, whereas \methname{} is entirely based on a dense urban deployment.

Finally, \methname{} provides both a raw dataset and a processed version in which CAM messages are interpolated to a fixed $10,Hz$ rate, with key kinematic attributes extracted and vehicle trajectories reconstructed, accounting for stationID changes introduced by ETSI privacy mechanisms. This additional processed release enables consistent time-series analysis and trajectory-level studies, whereas such an interpolated and pseudonym-reconciled version is not available in the V2AIX dataset.

A detailed comparison is provided through two complementary tables.~\Cref{tab:dataset_comparison_only_cits} presents a focused comparison between \methname{} and V2AIX, as the only two datasets based on standardized ETSI ITS communication messages. In addition,~\Cref{tab:dataset_comparison} provides a broader comparison including all the datasets discussed in the previous section, highlighting key characteristics such as the presence of CAM and DENM messages, the use of sensor-based data, the availability of real-life data, and the total distance covered by the recorded trajectories. This comparison illustrates that most existing datasets rely primarily on sensor data or simulated traces, whereas only a limited number provide real-world C-ITS communication logs. \methname{} offers the most extensive coverage in terms of traveled distance among the considered datasets.

\begin{table*}[t]
    \centering
    \renewcommand{\arraystretch}{1.0}
    \resizebox{\textwidth}{!}{%
    \begin{tabular}{lccccccc}
    \toprule
    &\multicolumn{2}{c}{\# of messages}&\multicolumn{2}{c}{Unique IDs}\\
    \textbf{Dataset} & CAM & DENM & CAM & DENM & \# of DENM Hazard Classes & \# of ITS message types & Frequency ($Hz$)\\
    \midrule
    \textbf{V2AIX} \cite{kueppers2024v2aix} & $263,467$ & $1338$ & $2.388$ & $37$ & \textbf{5} & \textbf{4} & --\\
    \textbf{\methname{} (ours)} & $\mathbf{40,446,000}$ & $\mathbf{2,376,571}$ & $\mathbf{25,839}$ & $\mathbf{1,332}$ & $3^*$ & 2$^*$ & $\mathbf{10}$\\
    \bottomrule
    \end{tabular}
    }
    \caption{Comparison table between ETSI-ITS-based datasets.}
    \vspace{5pt}
    \footnotesize{* numbers can be integrated with real-life data at any moment.}
    \label{tab:dataset_comparison_only_cits}
\end{table*}

\begin{table}[t]
    \centering
    \renewcommand{\arraystretch}{1.2} 
    \resizebox{\textwidth}{!}{%
    \begin{tabular}{lccccc}
    \toprule
    \textbf{Dataset} & CAM & DENM & \makecell[c]{Uses sensor \\ data} & \makecell[c]{Distance \\ covered ($Km$)} & \makecell[c]{Uses real-life \\ data} \\
    \midrule
    \textbf{Argoverse 2}~\cite{wilson2023argoverse} & \ding{55} & \ding{55} & \ding{51} & $2,220$ & \ding{51} \\
    \textbf{V2AIX} \cite{kueppers2024v2aix} & \ding{51} & \ding{51} & \ding{51} & $1,988$ & \ding{51} \\
    \textbf{PASMO}~\cite{ferreira2022dataset} & \ding{55} & \ding{55} & \ding{51} & -- & \ding{51} \\
    \textbf{Warrigal}~\cite{ward2014warrigal} & \ding{55}& \ding{55} & \ding{51} & 150 &  \ding{51}\\
    \textbf{TAPASCologne}~\cite{uppoor2013generation} & \ding{55}& \ding{55} & \ding{51} & -- &  \ding{55}\\
    \textbf{OPV2V}~\cite{xu2022opv2v} & \ding{55}& \ding{55} & \ding{51} & -- &  \ding{55}\\
    \textbf{Waymo Open}~\cite{sun2020scalability} & \ding{55} & \ding{55} & \ding{51} & -- & \ding{51} \\
    \textbf{ITD}~\cite{agarwal2024itd} & \ding{55}& \ding{55} & \ding{51} & -- &  \ding{51}\\
    \midrule
    \textbf{\methname{} (ours)} & \ding{51} & \ding{51} & \ding{55} & $14,287$ & \ding{51} \\
    \bottomrule
    \end{tabular}
    }
    \caption{Comparison table between the datasets.}
    \vspace{5pt}
    \label{tab:dataset_comparison}
\end{table}

\begin{figure}
    \subfloat[Distance and duration.]{
    \includegraphics[width=0.45\linewidth]{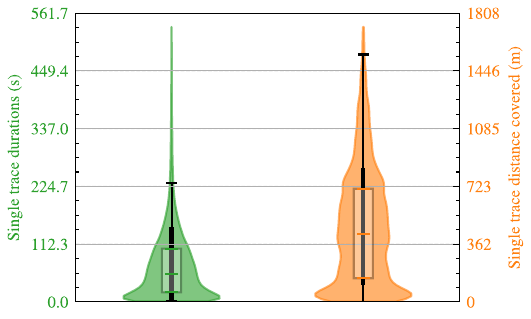}
    \label{fig:double_violin_StationID_switch_Speed(a)}
    }
    \hfill
    \subfloat[\% switch and avg speed.]{
    \includegraphics[width=0.45\linewidth]{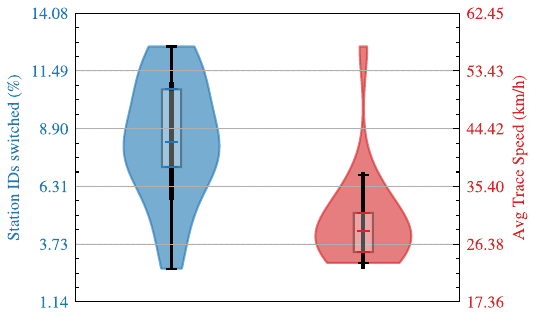}
    \label{fig:double_violin_StationID_switch_Speed(b)}
    }
    \caption{Dataset traces' statistics.}
    \label{fig:double_violin_StationID_switch_Speed}
\end{figure}

\nicepar{Statistical properties.}~
Some of the statistical properties of the processed dataset are illustrated in~\Cref{fig:double_violin_StationID_switch_Speed}, which reports the distributions of four key trajectory-level indicators across two dual-axis violin plots. \Cref{fig:double_violin_StationID_switch_Speed(a)} characterizes individual trace properties: the green distribution depicts trace duration (in seconds), while the orange distribution reports the distance covered per trace (in meters), both truncated at the 99th percentile to exclude extreme outliers. Both distributions exhibit a pronounced bottom skew, indicating that the vast majority of traces are short-lived and spatially compact, consistent with the limited but overlapping RSU coverage areas, whereas a non-negligible fraction of trajectories span considerably longer durations and distances, reflecting vehicles that traverse multiple RSU coverage zones in succession. 

\Cref{fig:double_violin_StationID_switch_Speed(b)} presents two complementary per-RSU indicators: the blue distribution captures the percentage of station ID switches observed within each RSU's reception range, while the red distribution reports the average speed of vehicles recorded per RSU. The switch rate distribution highlights significant variability across RSUs, reflecting heterogeneous traffic conditions and varying exposure to ETSI pseudonym-change events across the monitored urban area. The speed distribution, in contrast, is more concentrated, with values broadly consistent with urban speed limits, confirming that the dataset faithfully captures typical inner-city mobility patterns.

\nicepar{Limitations.}~
While \methname{} provides large-scale and longitudinal real-world V2X data, several limitations should be acknowledged.
First, data collection is entirely infrastructure-based and therefore spatially bounded by RSU deployment and radio coverage conditions. Consequently, trajectories are observable only within the monitored urban area and may be fragmented when vehicles leave coverage zones.
Second, the dataset reflects the penetration rate and technological characteristics of production vehicles equipped with ETSI-compliant V2X stacks in the MASA area. As such, mobility patterns and communication behaviors may be influenced by manufacturer-specific implementations and local traffic regulations.
Third, the pseudonym-reconciliation procedure, although empirically validated and constrained by physical motion consistency, remains a heuristic data-association process and may introduce residual false merges or splits in highly dense traffic scenarios.
Finally, \methname{} focuses on CAM and DENM messages and does not currently include additional ITS message types (\emph{e.g.}, SPATEM, MAPEM), which could further enrich intersection-level semantic context.
Despite these limitations, the dataset offers a statistically robust and operationally realistic foundation for C-ITS research in dense urban environments and can therefore support reproducible communication-aware mobility studies in real-world urban deployments.

\section{Conclusions and Future Works}
\label{sec_conclusions}
This paper introduced \methname{}, a large-scale infrastructure-based dataset derived from CAMs and DENMs collected within the MASA. With more than $40$ million CAMs and over $2$ million DENMs recorded under real urban traffic conditions, \methname{} represents one of the largest real-world ETSI ITS communication datasets currently available.

The dataset combines longitudinal continuity, high message density, and a pseudonym-reconciled trajectory-level release. The proposed reconciliation pipeline enables temporally consistent vehicle tracking despite ETSI privacy-driven stationID changes, while the $10Hz$ normalized version supports uniform time-series analysis and motion modeling. These characteristics provide a large-scale empirical basis for evaluating C-ITS solutions under realistic communication and mobility conditions.

With over $100,000 Km$ of reconstructed vehicle paths and a large population of unique station IDs, \methname{} offers representative urban mobility observations that can support not only trajectory prediction research, but also the calibration of microscopic traffic simulators (\emph{e.g.}, SUMO) using real V2X-derived kinematic traces. In addition, the dataset can contribute to the development of urban ITS Digital Twins by enabling joint modeling of traffic dynamics and V2X radio coverage conditions.

Future work will extend the dataset to include additional ITS message types and to enrich communication-centric analyses, including inter-arrival variability, multi-RSU reception overlap, and reliability indicators, to further characterize real-world C-ITS deployments.

\balance

\bibliographystyle{IEEEtran}
\bibliography{main}

\end{document}